\documentclass[notitlepage,superscriptaddress,showpacs,nobalancelastpage,twocolumn,aps,longbibliography,prb]{revtex4-2}
\usepackage[english]{babel}
\RequirePackage[T1]{fontenc}
\RequirePackage{times} 

\usepackage{siunitx} 
\usepackage[italicdiff]{physics}
\usepackage{amsfonts}
\usepackage{lipsum}
\usepackage{multirow}
\usepackage{subfigure}
\usepackage{amsmath}
\usepackage{amssymb}
\usepackage{graphicx, epstopdf}
\usepackage{wasysym}
\usepackage{xspace}
\usepackage{array}
\usepackage[hidelinks]{hyperref}
\usepackage[dvipsnames]{xcolor}
\usepackage{my_symbols}
\usepackage{CJK}
\usepackage{bm}
\usepackage{verbatim}
\usepackage{tikz}
\usepackage{comment}
\usepackage{marginnote}
\usepackage[textwidth=1.5cm]{todonotes}
\usepackage[normalem]{ulem}
\usepackage{soul}
\usepackage{booktabs}
\usepackage[commandnameprefix=ifneeded]{changes}

\sethlcolor{green}

{\par}

\newcounter{mycomment}

\excludecomment{comment}  

\begin{document}

\begin{CJK*}{UTF8}{gbsn} 
\title{Dynamic Control of Coupling Regimes in Oscillator Systems via Tunable Open Channels}
\author{Jiongjie Wang (王炯杰)}
\affiliation{Department of Physics and State Key Laboratory of Surface Physics, Fudan University, Shanghai 200433, China}
\author{Jiang Xiao (萧江)}
\email{xiaojiang@fudan.edu.cn}
\affiliation{Department of Physics and State Key Laboratory of Surface Physics, Fudan University, Shanghai 200433, China}
\affiliation{Institute for Nanoelectronics Devices and Quantum Computing, Fudan University, Shanghai 200433, China}
\affiliation{Shanghai Research Center for Quantum Sciences, Shanghai 201315, China}
\affiliation{Shanghai Branch, Hefei National Laboratory, Shanghai 201315, China}

\begin{abstract}
  We explore a system comprising two oscillators that are coupled to an open channel at distinct locations. The coupling nature can be adjusted to be coherent, dissipative, or a combination of both, controlled by a tunable phase resulting from wave propagation between the oscillators. This setup allows us to observe characteristic energy level behaviors: level repulsion occurs with coherent coupling, while level attraction is observed with dissipative coupling. In the regime of dissipative coupling, one of the eigenmodes becomes a dark mode, rendering it immune to external perturbations. By leveraging the tunable coupling through the open channel, we introduce a novel method for manipulating this dark mode, achieving both efficient excitation and an extended lifetime. Our results have broad applicability across various physical systems, including optical, acoustic, and magnetic configurations, underscoring the potential for innovative applications in mode storage and signal processing.
\end{abstract}

\maketitle
\end{CJK*}

\section{Introduction}

The study of coupled oscillators has long been essential for understanding complex phenomena across a range of physical systems, from condensed matter physics to quantum mechanics and engineering \cite{maleki_tunable_2004,karg_remote_2019,kahler_surface_2022,zhang_dissipative_2022}. Traditionally, research has concentrated on coherent coupling, a process characterized by a mutual exchange of energy that maintains phase relationships without dissipation. This often leads to the familiar anti-crossing pattern in energy spectra, a fundamental behavior in many physical systems that provides a robust framework for analyzing complex interactions \cite{aspelmeyer_cavity_2014,zare_rameshti_indirect_2018,li_strong_2019,an_coherent_2020,bae_exciton-coupled_2022,shen_coherent_2022,xu_quantum_2023,li_reconfigurable_2024,hou_coherent_2024}, 

In recent years, the concept of dissipative coupling has attracted significant attention as an alternative interaction mechanism, exhibiting a contrasting phenomenon known as level attraction \cite{yao_microscopic_2019,xiao_nonlinear_2021, wang_dissipative_2020,zare_rameshti_cavity_2022,grigoryan_pseudo-hermitian_2022, lu_synchronization_2023,qian_probing_2024,wang_enhancement_2024}. This interest is further amplified by the finding that coherent coupling can similarly induce level attraction within gain-loss systems \cite{eleuch_gain_2017, zhang_exceptional_2024,zhang_gain-loss_2024}. Dissipative coupling is particularly captivating due to its capacity to produce exceptional points, where eigenvalues and eigenvectors converge, resulting in non-Hermitian degeneracies \cite{el-ganainy_non-hermitian_2018,wang_non-hermitian_2023}. Researchers have proposed various mechanisms to implement dissipative coupling, including non-reciprocal coupling that leverages asymmetric interactions to preferentially direct transmission \cite{metelmann_nonreciprocal_2015}, and external phase delay lines that allow for precise modulation of interaction phases \cite{grigoryan_synchronized_2018, rao_braiding_2024}. Additionally, intermediate modes characterized by considerable damping can effectively mediate interactions with phase delay \cite{yang_anti-_2017, yu_prediction_2019}, enhancing overall control. These diverse and adaptable mechanisms highlight the potential of dissipative coupling, paving the way for further exploration and innovative applications in various fields of study.


In the regime of dissipative coupling, the presence of dark modes-eigenmodes characterized by minimal dissipation-presents intriguing opportunities for advanced applications. These dark modes, which are largely immune to external perturbations, pose significant challenges for excitation, in contrast to bright modes that, while easily excitable, experience rapid decay. Notably, bound states in the continuum (BICs), which can be protected by symmetry \cite{hsu_bound_2016} or topology \cite{zhen_topological_2014}, exemplify a class of dark modes that remain decoupled from their environment. For instance, in a double-oscillator system coupled to an open channel, the anti-symmetric mode serves as a dark mode that can be accessed by breaking the system's inherent symmetry \cite{plotnik_experimental_2011, gentry_dark_2014} or by introducing dissipation into the channel \cite{sadrieva_transition_2017, muhammad_radiationless_2024}. Furthermore, accidental BICs have been identified \cite{koshelev_nonradiating_2019}, and research indicates that nonlinear interactions between two oscillators can facilitate the emergence of dark modes, thereby enabling efficient energy injection \cite{bulgakov_bound_2010, bulgakov_all-optical_2015}. Techniques such as adjusting the spacing between microwave cavities \cite{panaro_dark_2014} or utilizing quantum tunneling \cite{feist_extraordinary_2015} have been explored to convert dark modes into bright ones. However, these manipulations often compromise the uncoupling properties of dark modes or limit their excitation efficiency. 

In this paper, we explore the realization of coherent and dissipative coupling between two oscillators positioned at different locations along an open channel. The coupling regime can be effectively modulated by adjusting the separation distance between the oscillators or by tuning their resonant frequencies. This methodology significantly enhances the flexibility of coupling dynamics, facilitating precise control over both the coupling regime and mode manipulation. Notably, in the dissipative coupling regime, we observe the presence of a dark mode that remains decoupled from the external environment via the open channel. Our findings reveal two distinct dissipative regimes characterized by an additional $\pi$ phase shift relative to one another, leading to a swapping of dark and bright mode statuses between these regimes. This dynamic interchangeability enables the efficient excitation of modes while simultaneously extending their lifetimes, thereby presenting new opportunities for advanced applications in mode or energy injection and storage.

This paper is organized as the following: Section \ref{sec:model} presents the model employed to depict the channel-mediated coupling modes and examines the eigenmodes' characteristics across varying coupling types. Section \ref{sec:real} demonstrates the practical realization of modulating coherent and dissipative coupling within three specific systems: optical, acoustic, and magnetic. Finally, Section \ref{sec:dark} explicates the method of achieving both simultaneous excitation and extended mode storage by switching between divergent dissipative coupling regimes.

\section{Modes Coupled via Open Channel}
\label{sec:model}

\begin{figure}[t]
  \includegraphics[width=0.98\columnwidth]{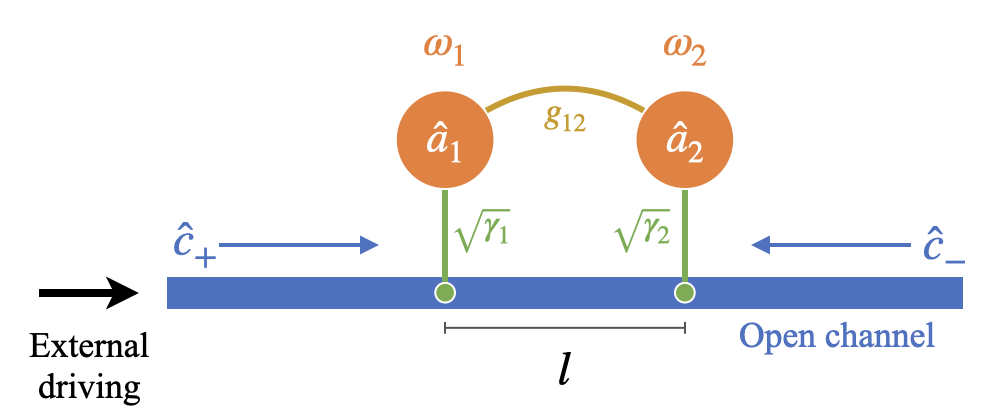}
  \caption{Two coupled oscillator in contact with an open channel. The open channel may introduce additional dissipation as well as additional coupling for the two oscillators. The extra coupling is in general non-reciprocal because of the finite separation $l$ between the two oscillators in space introduce an non-zero phase for the coupling.}
  \label{fig:model}
\end{figure}

We consider two coupled modes in contact with an open channel as depicted in \Figure{fig:model}. The system Hamiltonian
\begin{equation}
  \label{eqn:Ham}
  \hH = \hH_a + \hH_c + \hH_{ac}
\end{equation}
consists of i) the contribution from the two modes $a_{1,2}$ with intrinsic frequency $\omega_{1,2}$ that are coupled via $g_{ij}$: 
\begin{equation}
  \label{eqn:Ha}
  \hH_a = \hbar\omega_1 \ha_1^\dagger\ha_1 +\hbar\omega_2 \ha_2^\dagger\ha_2 + \hbar(g_{12} \ha_1^\dagger\ha_2 + g_{21} \ha_1 \ha_2^\dagger); 
\end{equation}
ii) the contribution from the open channel with left- and right-gong propagating modes $\hc_\pm(\omega)$: 
\begin{equation}
  \label{eqn:Hc}
  \hH_c = \int \dd{\omega} \hbar\omega \qty[\hc_+^\dagger(\omega)\hc_+(\omega) + \hc_-^\dagger(\omega)\hc_-(\omega)];
\end{equation}
and iii) the coupling between the modes and the open channel: 
\begin{equation}
  \label{eqn:Hac}
  \hH_{ac} = \sum_{i=1,2}^{\chi = \pm}
  \int {\frac{\hbar }{\sqrt{2\pi}}} 
  d\omega \qty[\sqrt{\gamma_i}e^{i\chi \abs{q}x_i}\ha_i^\dagger \hc_\chi(\omega) + \mbox{c.c.}],
\end{equation}
where $\sqrt{\gamma_i}$ is the coupling strength between the mode-$i$ and the open channel, and an additional phase $qx_i$ appears because the oscillator-$i$ is in contact with the channel at position $x_i$. Here $q$ is wavevector of the wave propagating in the channel at frequency $\omega$. 

By integrating out the propagating modes $\hc_\pm$, we obtain the Heisenberg equation of motion for $\ha_{1,2}$ as
\begin{equation}
  \label{eqn:eom}
  -i\dv{t}\mqty(\ha_1 \\\ha_2) =
  \mqty(
    \omega_1 - i|\gamma_1| & \kappa_{12} \\
    \kappa_{21} & \omega_2 - i|\gamma_2|)
    \mqty(\ha_1 \\\ha_2),
\end{equation}
where $\kappa_{ij} = g_{ij} - i\sqrt{\gamma_i^*\gamma_j}e^{i\abs{ql}}$ with $l = \abs{x_2-x_1}$ includes the direct coupling between the two modes and the coupling mediated by the open channel. 

For the rest of the paper, we assume that the two modes are not directly coupled, \ie $g_{21} = g_{21} = 0$. To simplify the symbols further, we assume $\kappa = \sqrt{\gamma_1^*\gamma_2} = \sqrt{\gamma_2^*\gamma_1} = \abs{\gamma_1} = \abs{\gamma_2}$. Hence, the effective Hamiltonian for the equation of motion \Eq{eqn:eom} can be rewrite as 
\begin{equation}
  \label{eqn:Heff}
  H_\text{eff} =  \mqty(
    \omega_1 - i\kappa & \kappa e^{i\theta} \\
    \kappa e^{i\theta} & \omega_2 - i\kappa),
\end{equation}
where $\kappa$ and $\theta$ represents the strength and phase of the channel-mediated coupling, respectively. 
The $i\kappa$ in the diagonal elements is the additional dissipation introduced due to leakage via the open channel.
The phase factor depending on the separation of the two oscillators along the channel $\theta = \abs{ql} - \pi/2$.
It is noteworthy that this phase is identical for two off-diagonal terms, indicating that the channel-mediated coupling is inherently non-reciprocal or, more generally, non-Hermitian when \(\theta \neq n\pi\). 
This non-reciprocity arises from the delay due to propagation from one mode on the other via the open channel.
Furthermore, when $ie^{i\abs{ql}}$ is real (imaginary), the channel-mediated coupling is a coherent (dissipative) type. Since this phase can be tuned by varying the separation $l$ or the wavevector $q$, it is possible to tune the nature of the coupling between the two modes, from coherent to dissipative or vice versa. It is this tunability in the coupling and its applications that we want to explore in this paper.

\begin{figure}[t]
  \includegraphics[width=0.98\columnwidth]{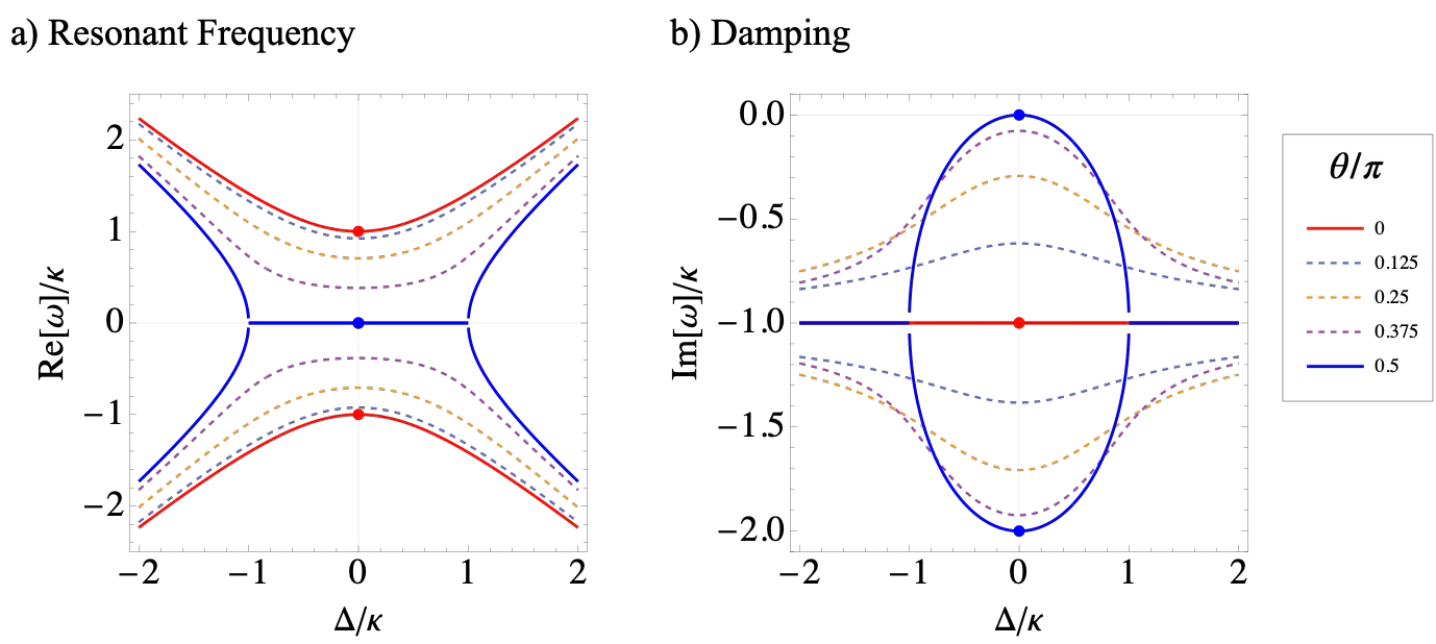}
  \caption{The real (a) and imaginary (b) part of the eigenfrequencies as function of the detuning $\Delta$ for for various coupling phase, the level repulsion (red) for $\theta = 0$ and level attraction (blue) for $\theta = \pi/2$.
  }
  \label{fig:LALR}
\end{figure}

\Figure{fig:LALR} depicts the real and imaginary part of the spectrum \Eq{eqn:Heff} as function of the detuning $\Delta = \Re{\omega_1 - \omega_2}$ for different values of the coupling phase $\theta$. As $\theta = 0$, the coupling is coherent, and the spectrum shows the standard level repulsive behavior, and the dissipation for the two normal modes are the same, corresponding to the solid red curves in \Figure{fig:LALR}(a) and (b) respectively. On the other hand, when $\theta = \pi/2$, the coupling is purely dissipative, and the spectrum shows the level attractive behavior, corresponding to the solid blue curves in \Figure{fig:LALR}(a) and (b) respectively. In the level attractive case, the two normal modes show very different features in terms of dissipation, one mode has reduced dissipation and longer lifetime, while the other has enhanced dissipation and shorter lifetime. 


It is crucial to recognize that the spectra for the system depicted in \Figure{fig:model} cannot be accurately measured using the standard \( S_{21} \) transmission through the open channel. Previous study by the authors has shown that \( S_{21} \) transmission spectra may yield inaccurate spectrum when the detection channel mediates extra coupling \cite{wang_interpreting_2024}. Because the open channel in this study actually mediates such coupling, the \( S_{21} \) measurements obtained via the same open channel will not represent the true spectra. Therefore, it is essential to utilize either the energy spectra or the \( S_{21} \) spectra measured through an independent detection channel that does not permit coupling.


\begin{center}
  \label{tab:gg}
  \begin{table}[t]
  \begin{tabular}{p{2cm}||p{1.4cm}|p{1.4cm}||p{1.4cm}|p{1.4cm}} 
  Coupling type & \multicolumn{2}{c||}{Coherent} & \multicolumn{2}{c}{Dissipative} \\ \hline
  Coupling phase & \multicolumn{2}{c||}{$\theta = n\pi$} & \multicolumn{2}{c}{$\theta = n\pi+\pi/2$} \\ \hline
  Spectrum & \multicolumn{2}{c||}{Level Repulsion (LR)} & \multicolumn{2}{c}{Level Attraction (LA)} \\ \hline
  $n$ & even & odd & even & odd \\ \hline
  in-phase mode & low frequency & high frequency & low damping & high damping \\ 
  & {\it bright} & {\it bright} & {\it dark} & {\it bright} \\ \hline
  out-of-phase mode & high frequency & low frequency & high damping & low damping \\ 
  & {\it bright} & {\it bright} & {\it bright} & {\it dark} \\ \hline
  & LR$^+$ & LR$^-$ & LA$^+$ & LA$^-$ 
  \end{tabular}
  \caption{The correspondence between the coupling type and spectrum features and the eigenmodes symmetry.  }
 \end{table}
\end{center}

Apart from drastic different spectra in \Figure{fig:LALR}, the coherent and dissipative coupled systems also manifest distinct characters of eigenmodes, in terms of their darkness or brightness in particular. The classification of internal modes into bright and dark categories is essential for understanding the dynamics of coupled physical systems \cite{lin_coherent_2010, dong_optomechanical_2012}. Bright modes are highly responsive to external perturbations, exhibiting strong coupling with their environment, which results in significant energy dissipation and shorter lifetimes. In contrast, dark modes demonstrate a remarkable resilience to external influences, characterized by weak coupling and minimal energy loss, leading to longer lifetimes. This distinction is particularly evident in dissipatively coupled systems, such as the open channel model, where both modes share identical resonance frequencies but differ markedly in their excitation efficiencies and damping characteristics. 

The eigenmodes of coupled-oscillator model are usually labeled as in-phase and out-of-phase modes, or sometimes referred as the acoustic and optical modes. For the coherent coupling cases (coupling phase $\theta = n\pi$), it is generally true that the out-of-phase mode has higher frequency than the in-phase mode in most physical scenarios, which has the coupling phase $\theta = 0$. However, when the coupling phase $\theta = \pi$, the out-of-phase mode would have lower frequency than the in-phase mode. Such behavior can be generalized as in Table I that the in-phase (out-of-phase) mode has lower (higher) eigenfrequency when $\theta = n\pi$ for even $n$, and the opposite for odd $n$, which we label as the LR$^+$ and LR$^-$, respectively. 

For the dissipatively coupled case with coupling phase $\theta = n\pi + \pi/2$, the difference between the in-phase and out-of-phase modes are not in their resonant frequencies, but in their damping. For even $n$, the in-phase (out-of-phase) mode has low (high) damping, and the opposite for odd $n$, which we label as LA$^+$ and LA$^-$, respectively. Since the dissipation we considered here is due to the coupling with the open channel, the high (low) damping mode is the bright (dark) mode, which means the in-phase (out-of-phase) mode can be either bright or dark depending on the evenness of $n$ in the coupling phase. This $n$-dependence of the brightness (darkness) enables us to alter the brightness or darkness of a certain mode.

\section{Realization in Physical Systems}
\label{sec:real}

In this section, we show the realization of the channel-mediated non-reciprocal coupling based on the optical system, acoustic system, and magnetic system. In all case, the nature of the coupling can be effectively tuned by varying the coupling phase. 

\begin{figure*}[t]
  \centering
  \includegraphics[width=\textwidth]{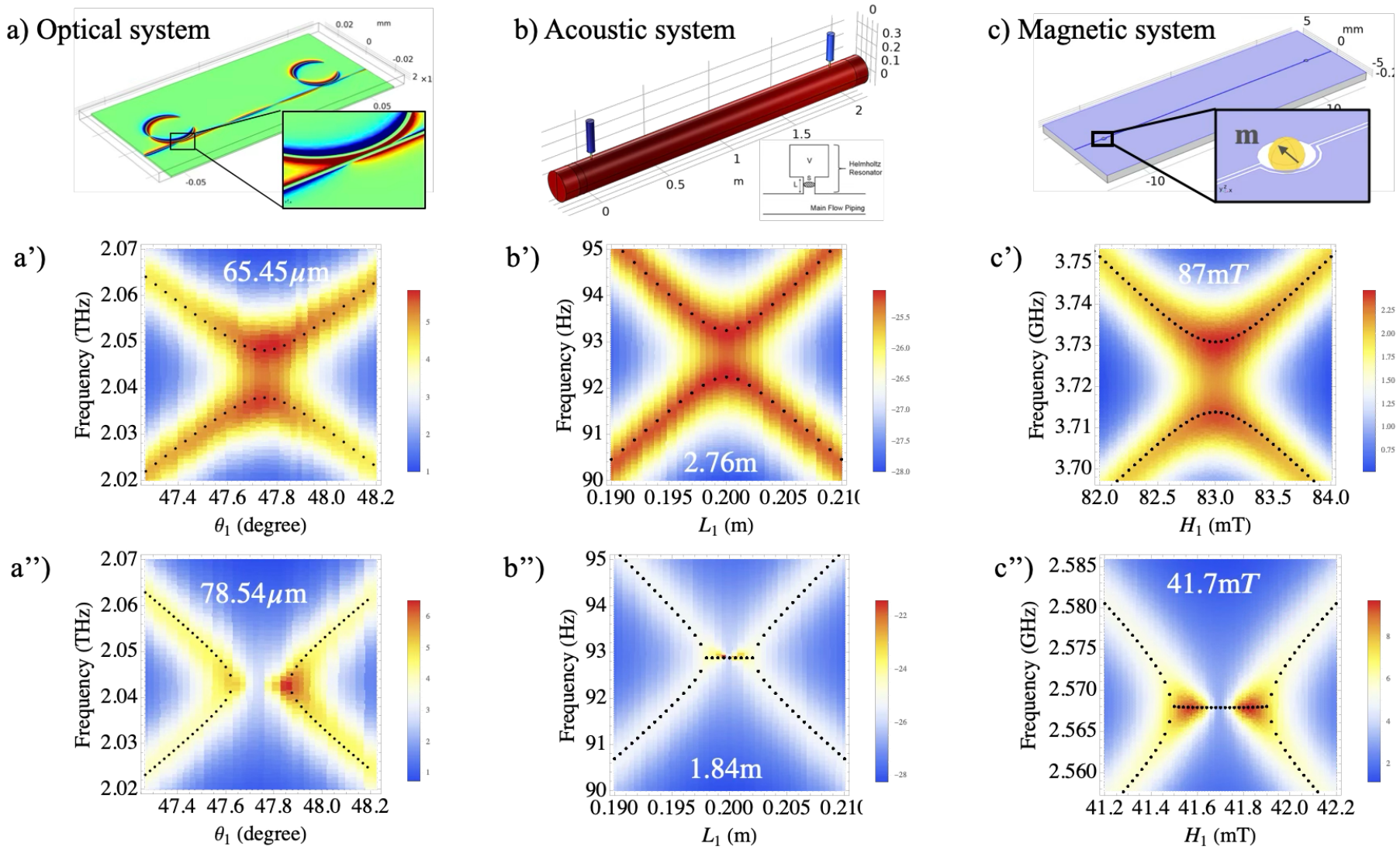}
  \caption{
    a). Optical system: Two C-shaped micro-ring cavities coupled via a microstrip in proximity. The color map shows a typical magnetic field distribution perpendicular to the film. 
    a'). A level repulsive (LR) spectrum in the excitation energy for \( l = \SI{78.54}{\micro\metre}, \theta = 2.5\pi \). 
    a'') A level attraction (LA) spectrum for \( l = \SI{65.45}{\micro\metre}, \theta = 2\pi \). 
    b). Acoustic system:  Two Helmholtz resonators coupled via an open tube. 
    b'). LR spectrum at $l = \SI{276}{\micro\metre}, \theta = \pi$. 
    b''). LA spectrum at $l = \SI{184}{\micro\metre}, \theta = 0.5\pi$. 
    c). Magnetic system: two YIG spheres coupled via a co-planar waveguide. 
    c'). LR spectrum at $H_0 = \SI{87}{\milli\tesla}, \theta = 2\pi$. 
    c''). LA spectrum at $H_0 = \SI{41.7}{\milli\tesla}, \theta = 1.5\pi$.
    }
  \label{fig:LALR_demo}
\end{figure*}

\subsection{Optical system}

The optical system consists of two C-shaped micro-ring resonators positioned beside a microstrip line, forming an integrated photonic circuit \cite{schiattarella_terahertz_2024}. Each micro-ring, typically made from high-refractive-index materials like silicon, traps light through total internal reflection, while the microstrip line, a conductive strip on a dielectric substrate, guides microwave signals. The proximity of the resonators to the microstrip line allows for evanescent coupling, whose strength is controlled by the spacing between the C-ring and the microstrip. The Hamiltonian given in \Eqss{eqn:Ham}{eqn:Hac} can describe this system perfectly, with $\ha_{1,2}$ being the resonant modes in the C-shaped micro-ring, and $\hc_\pm(\omega)$ being the propagating waves in the microstrip. 

The geometric and material parameters are as the following:
the C-shaped micro-ring has an opening angle of $\pi/3$ and radius \SI{10}{\mu m}, and the widths of the micro-ring and microstrip line are both \SI{0.5}{\mu m}. The edge-to-edge spacing between the micro-ring and the microstrip line is \SI{0.5}{\mu m}. The two rings are separated by a distance $d \sim 50 - \SI{100}{\mu m}$ along the microstrip. The dielectric constant of the silicon substrate is 11.68. According to these parameters, the resonant frequency of the two micro-ring is $\omega \sim \SI{2.04}{THz}$, 
and the corresponding wavelength of the electromagnetic wave in the microstrip at this frequency is $\lambda \simeq \SI{52.36}{\mu m}$. 

At this wavelength, the two rings are coherently coupled for $d \simeq 1.25\lambda = \SI{65.45}{\mu m}$, and dissipatively coupled for $d \simeq 1.5\lambda = \SI{78.54}{\mu m}$, corresponding to a delay phase $\theta = 2\pi$ and $\theta = 5/2\pi$, respectively. 
We artificially tune the frequencies of the two microrings by varying the opening angle of the C-ring.
The simulated spectra in COMSOL Multiphysics confirm the expected results: the system exhibits the level repulsion spectrum for $d = \SI{65.45}{\mu m}$ as in \Figure{fig:LALR_demo}(a').
and the level attraction for $d = \SI{78.54}{\mu m}$ as in \Figure{fig:LALR_demo}(a''). These patterns are in agreement with \Figure{fig:LALR}(a) for the cases with the corresponding coupling phases. 
It should be noted that for the level repulsion spectrum, the gap opened by the channel-mediated coupling is comparable to the bordering due to the open channel, therefore the double peak is nearly closed.



\subsection{Acoustic System}

Similar effects can also be demonstrated using acoustic systems, exemplified by the Helmholtz resonator functioning as an acoustic oscillator \cite{long_6_2014}, as depicted in the inset of \Figure{fig:LALR_demo}(b). This device comprises an acoustic cavity of volume \( V \) connected to the atmosphere through a neck of length \( L \) and cross-section \( S \). The eigenfrequency of the Helmholtz resonator is expressed by the formula \( \omega = c \sqrt{S/LV} \) \cite{long_6_2014}, where \( c = \SI{343}{cm/s} \) denotes the speed of sound in air. In this configuration, the air within the neck and cavity acts analogously to the mass and spring components of a mechanical oscillator, respectively. 



We now consider two of such Helmholtz resonators connected an open tube with cross-section $S_0$ at different locations as illustrated in \Figure{fig:LALR_demo}(b). Here the open tube serves as the coupling channel.
The effective Hamiltonian for the equation of motion for this acoustic system is the same as \Eq{eqn:Heff}, but with coupling strength and coupling phase given by (see Appendix \ref{app:acoustic})
\begin{equation}
  \kappa_\text{ac} = \frac{Sc}{4S_0L}
  \qand
  \theta = ql - \frac{\pi}{2},
\end{equation}
where $q = \omega/c = 1/\lambda$ is the wavevector  corresponding to the frequency $\omega$.

The parameters of the acoustic system are as below: 
$S = \SI{0.785}{cm^2}, L = \SI{6.875}{cm}, V_i = \SI{19.625}{cm^2}\times h_i, S_0 = \SI{314}{cm^2}$.
The resonant frequencies of the individual resonator is varied via the height of the cavities:
$h_{1,2} = 19 \sim \SI{21}{cm}$ while keeping $h_1 + h_2 = \SI{40}{cm}$ constant, so that the frequency of one oscillator increases and the frequency of another oscillator decreases. 
With these parameters, the resonance frequency of the Helmholtz resonator is about \SI{93}{Hz} for $h_{1,2} = \SI{20}{cm}$, corresponding to a wavelength of $\lambda = \SI{369}{cm}$.

\Figure{fig:LALR_demo}(b',b'') show the simulated resonance spectra in COMSOL Multiphysics for $l = \SI{184}{cm} \simeq 0.5\lambda$ and $\SI{276}{cm} \simeq 0.75\lambda$, corresponding a coupling phase of $\pi/2$ and $\pi$, or the coherent coupling and dissipative coupling, respectively. Both spectra are in good agreement with the theoretical expectations. 
A special signature for the LA spectrum at the zero detuning point shows an exceptionally wide linewidth. 
This is because at this point the small dissipation mode is decoupled from the outside world, and is not excited at all, leaving only the large dissipation mode with a wider linewidth.
Similar to the optical case, here we read the sum of the energies in the two Helmholtz resonators, rather than the $S_{21}$ spectrum, to demonstrate the resonant characteristics of the system.

\subsection{Magnetic System}

We now direct our focus to the magnetic configuration, which comprises two yttrium iron garnet (YIG) spheres with low intrinsic damping, placed adjacently to a coplanar waveguide with certain separation from one another. Each sphere operates as an individual oscillator utilizing the macrospin precessional mode, whose resonant frequency can be precisely adjusted through the application of an external magnetic field. 
Similar to the optical and acoustic scenarios, the effective Hamiltonian for the equation of motion governing the macrospin excitation amplitudes in the two YIG sphere is the same as \Eq{eqn:Heff} with
coupling strength and coupling phase given by (see Appendix \ref{app:magnetic})
\begin{equation}
  \kappa_\text{mag} = \frac{qd\omega_M}{4}
  \qand
  \theta = ql + \frac{\pi}{2}.
\end{equation}

The parameters of the system are: the width of the central metal line in the coplanar waveguide and the dielectric regions on both sides are both \SI{20}{\mu m}, the radius of the semicircle surrounding the YIG is \SI{0.2}{mm}, and the relative permittivity of the dielectric material beneath the coplanar waveguide is 11.68 with a thickness of \SI{0.762}{mm}. Each YIG sphere has a radius of \SI{0.125}{mm} and is placed at the center of the semicircle formed by the coplanar waveguide. 
The two YIG spheres possess an easy-axis anisotropy field of \SI{48.76}{mT}\cite{yan_all-magnonic_2011}, corresponding to a resonance frequency of $\omega_0 = \SI{8.58}{GHz}$. 
The separation between them along the waveguide $l = \SI{22.9}{mm}$. External magnetic field can be used to tune the resonant frequencies of the YIG sphere: the uniform field is to shift the average frequency, thus the coupling phase, and a field gradient across the two spheres determines the detuning.

In the simulation, we only focus on the coupling of the uniform precession modes, whose frequency depends on the external field $H_0$: $\omega\simeq \SI{8.58}{GHz} + H_0\times\SI{0.176}{GHz/mT}$. At the external magnetic field $H_0 = \SI{41.7}{mT}$, the system's frequency $\omega \simeq \SI{15.9}{GHz}$ and the wavelength $\lambda \simeq \SI{45.8}{mT}$. At $H_0 = \SI{87}{mT}$, the system's frequency $\omega \simeq \SI{23.9}{GHz}$ and the wavelength $\lambda \simeq \SI{30.5}{mm}$.

Varying the separation between the two YIG sphere is difficult in practice. It is much easier to change the resonant frequency by varying the external magnetic field, thus realizing the switching between level repulsive ($H_0 = \SI{87}{mT}, \theta = 2\pi$) and attractive ($H_0 = \SI{41.7}{mT}, \theta = 3\pi/2$), as simulated by COMSOL Multiphysics in \Figure{fig:LALR_demo}(c',c''). 


\section{Excitation and Long-Term Storage of Dark Mode via Tunable Coupling}
\label{sec:dark}

\begin{figure*}[t]
  \centering
  \includegraphics[width=0.98\textwidth]{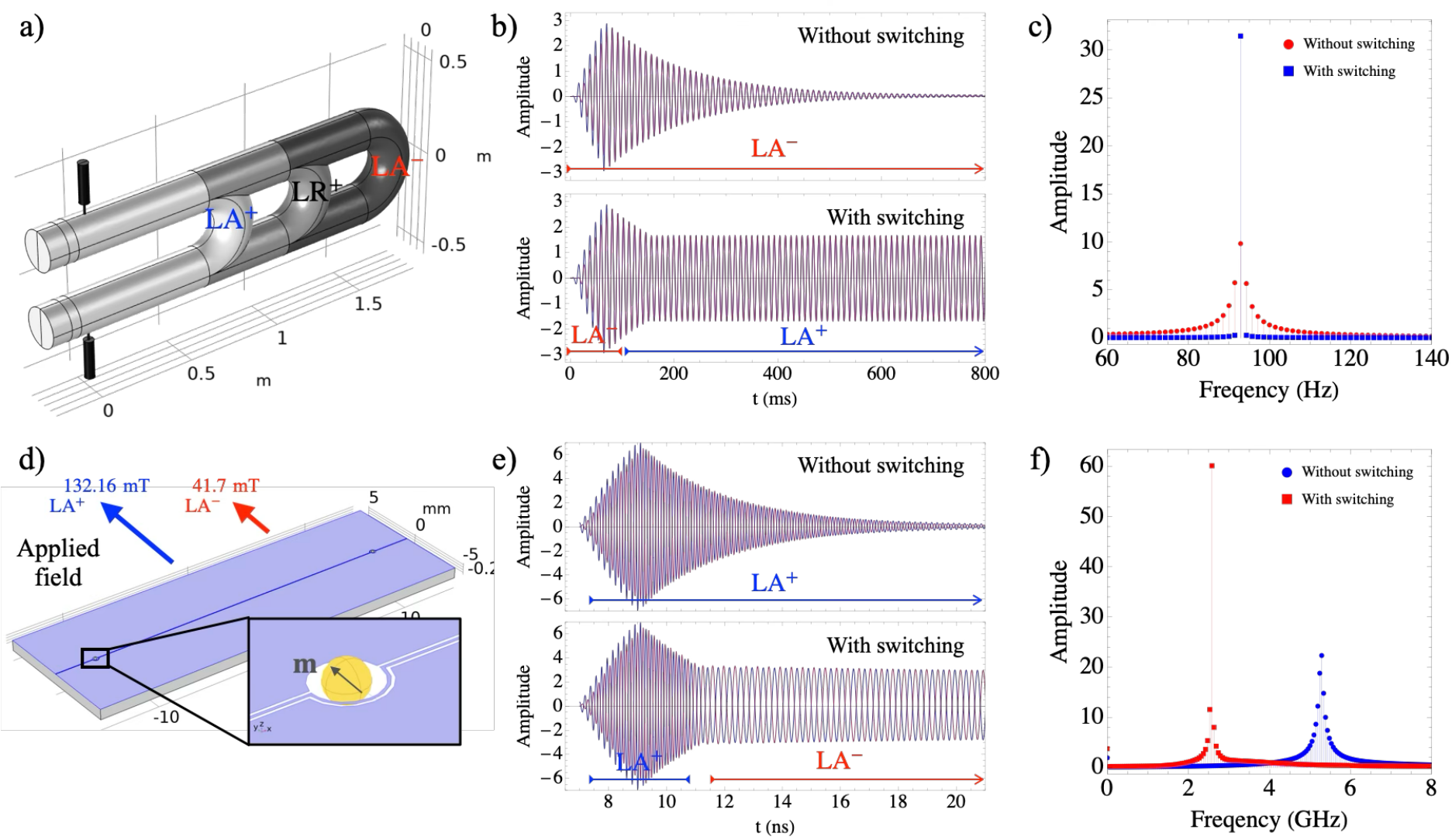}
  \caption{
    a). The acoustic system with three connecting routes realizing LA$^-$, LR$^+$, and LA$^+$, respectively.
    b). Upper: The amplitudes of the average pressure in the cavity of the Helmholtz resonators as function of time with the coupling via route LA$^-$ throughout. The excited in-phase mode is quickly damped out via the open channel. 
    Lower: Same as the upper panel but with a configuration switching from LA$^-$ to LA$^+$ at $t = \SI{100}{ms}$. The in-phase mode excited during the LA$^-$ phase becomes dark after switching, thus amplitude sustains much longer.
    d). The magnetic system with configuration switching from LA$^+$ to LA$^-$ by reducing the external field from \SI{132.16}{mT} to \SI{41.7}{mT}.
    c). The Fourier transform of b) for the mode oscillation with and without LA$^-$ to LA$^+$ switching.
    e). Upper: The amplitudes of $m_z^{1,2}$ as function of time with fixed external field $\SI{132.16}{mT}$ (LA$^+$ configuration) throughout. The excited out-of-phase mode is quickly damped out via the open channel. 
    Lower: Same as the upper panel but with a configuration switching from LA$^+$ to LA$^-$ by gradually reducing the external field from \SI{41.7}{mT} around $t \simeq \SI{11}{ns}$. The out-of-phase mode is efficiently excited during the LA$^+$ phase now becomes dark after switching, thus amplitude sustains much longer.  
    f). The Fourier transform e) for the mode oscillation with and without LA$^+$ to LA$^-$ switching.
    }
  \label{fig:LALA}
\end{figure*}

As illustrated in Table \ref{tab:gg}, the brightness or darkness of specific mode depends critically on the exact coupling phase in the scenario of dissipative coupling. Notably, the in-phase mode exhibits a dark state when the coupling phase $\theta = \pi/2$ for even integers \(n\) (LA$^+$), while it transitions to a bright state for odd integers \(n\) (LA$^-$). Conversely, the out-of-phase mode behaves in the opposite manner, demonstrating brightness for even \(n\) and darkness for odd \(n\). This dependency underlines the potential for tuning the visibility of particular modes by adjusting the coupling phase. Such a capability presents a strategy for simultaneous optimization of both efficient excitation and long-term storage.

The process under discussion consists of two distinct steps. Initially, the system is configured in the LA$^-$ setting with the coupling phase set to \(\theta = 3\pi/2\), which renders the in-phase mode a bright mode (see Table \ref{tab:gg}). This configuration permits efficient excitation of the mode via an external driving field transmitted through the open channel. Once the in-phase mode achieves full excitation, the external driving is turned off, and the system is reconfigured to the LA$^+$ configuration by adjusting the coupling phase to \(\theta = \pi/2\). In this new state, the in-phase mode transitions to a dark mode, leading to the elimination of the channel-induced dissipation and thereby extending its operational lifetime. This methodology allows for the simultaneous efficient injection of the in-phase mode in the LA$^-$ configuration and its long-term preservation in the LA$^+$ configuration. A similar process can be designed for the out-of-phase mode by reversing the order of the LA$^+$ and LA$^-$ configurations.

We first demonstrate the two-step process within an acoustic system designed to achieve an in-phase mode with an extended lifetime. Figure \ref{fig:LALA}(a) illustrates the experimental setup, which comprises two Helmholtz resonators with identical intrinsic frequencies of $\omega_1 = \omega_2 = \SI{93}{Hz}$, interconnected via three uniquely configured paths.
The lengths of these routes are specifically chosen to have the coupling phase match the LA$^+$ ($l_1 = \SI{184}{cm}, \theta = \pi/2$), LR$^-$ ($l_2 = \SI{277.5}{cm}, \theta = \pi$), and LA$^-$ ($l_3 = \SI{371}{cm}, \theta = 3\pi/2$), respectively. Initially, the system is configured along route 3, establishing a coupling phase of $\theta = 3\pi/2$, thereby in the LA$^-$ configuration. An acoustic excitation is applied through one open end of the tube for $\SI{64.3}{ms}$, effectively exciting the in-phase (bright) mode. 
We explore two scenarios: first, keeping the system on route 3 after the driving is turned off at $t = \SI{64.3}{ms}$, and second, switching the system to route 1 (LA$^+$) at $t = \SI{150}{ms}$ following the field cut-off. 
In the first scenario, the bright in-phase mode suffers rapid amplitude decay due to dissipation through the open channel throughout, acquiring a lifetime of approximately $\SI{200}{ms}$, as depicted in Figure \ref{fig:LALA}(b). Conversely, in the second scenario, although some decay occurs post-excitation pre-switch, the mode transitions into a dark mode upon route-switching, significantly reducing the decay rate and extending the mode's lifetime nearly indefinitely, limited only by the intrinsic dissipation. \Figure{fig:LALA}(c) shows the Fourier transform of the wave forms without and with switching, showing extremely sharp peak for the case with switching.

The phenomenon of dark mode excitation and long-term retention can also be demonstrated in magnetic systems. An advantage in magnetic systems is the simplicity in switching between LA$^+$ and LA$^-$ configurations - by merely adjusting the external magnetic field. This adjustment alters the resonance frequencies of the (YIG) sphere, affecting the wavevector \( q \) and hence the coupling phase. The subfigure at the top in \Figure{fig:LALA}(e) illustrates the excitation and damping of the out-of-phase (bright) mode when the system is held at the LA$^+$ configuration throughout the process. Meanwhile, the subfigure at the bottom demonstrates an excitation at the LA$^+$ configuration with an external field \( H_0 = \SI{137}{mT} \) prior to \( t = \SI{11}{ns} \). A switch to the LA$^-$ configuration occurs after that as the external field is reduced to \( H_0 = \SI{41.7}{mT} \). Post-switching, the out-of-phase mode transitions to a dark state, exhibiting prolonged retention with minimal dissipation. Nevertheless, even in this low-dissipation mode, a certain degree of energy loss persists due to the YIG spheres' coupling with the surrounding dielectric medium and the intrinsic Gilbert damping of the material. 
\Figure{fig:LALA}(f) shows the Fourier transform of the wave forms without and with switching, showing extremely sharp peak for the case with switching.

\section{Discussion \& conclusions}

In order to realize meaningful switching between LA$^+$ and LA$^-$ configurations, we require that the switching time $\tau$ should be short such that the energy dissipated (characterized by coupling strength to the open channel $\kappa$) during the switching is negligible, \ie $\tau \ll 1/\kappa$. 
On the other hand, the switching time should be long enough so that the switching act does not influence the system. In other words, the switching should be quasi-static, or $\tau \gg 1/\omega_{1,2}$. Overall, we require
\begin{equation}
  \kappa \ll \tau^{-1} \ll \omega_{1,2}.
\end{equation}


\Figure{fig:LALA} illustrates that the transition from LA$^-$ to LA$^+$ results in the swapping of bright and dark modes, and realizes efficient excitation and long-term storage of one eigenmode.
It is also possible to tune the coupling to realize switching between LR and LA configurations, such as LR$^+$-LA$^+$ switching.
In this scenario, both the in-phase and out-of-phase modes are bright and can be effectively excited. Upon switching to the LA$^+$ configuration, the out-of-phase mode retains its brightness while the in-phase mode becomes dark. As a result, the out-of-phase mode experiences rapid decay, whereas the in-phase mode is sustained over an extended period.


In summary, we have demonstrated that an open channel can be used to couple two harmonic subsystems, and the nature of the coupling can be tuned continuously from coherent to dissipative. The physical principle is general and applicable to almost all types of resonators, including the optical, acoustic, and magnetic resonances.
The utilization of tunable coupling represents a strategy in mode activation and storage. By managing the coupling conditions, it is possible to achieve efficient excitation and long-term preservation of modes simultaneously, overcoming the traditional barriers posed by dissipation. 



%

\section{Acknowledgements}

This work was supported by the National Natural Science Foundation of China under Grant No. 12474110, the National Key Research and Development Program of China (Grant No. 2022YFA1403300) and Shanghai Municipal Science and Technology Major Project (Grant No. 2019SHZDZX01). 

\bibliographystyle{apsrev4-2}
\bibliography{ref_tune_coupling}

\bigskip
\appendix

\section{Acoustic System}
\label{app:acoustic}

The Lagrangian of the acoustic system is:
\begin{align}
  \cL &= \half \sum_{i=1,2} \rho SL\qty(\dot{\xi}_i^2 - \omega_i^2 \xi_i^2)
  - \sum_{i=1,2} p(x_i)S\xi_i \nn
  &+ \half\int dx S_0\rho\qty[v(x)^2 - \frac{1}{\rho^2c^2} p(x)^2],
\end{align}
where $\omega_{1,2} = c\sqrt{S/LV_{1,2}}$ are the resonant frequencies of the two individual Helmholtz resonators, $c$ is the speed of sound in air, $S$ and $L$ are the cross-section and length of the neck of resonator, $V_{1,2}$ are the volume of the head of the two resonators, $S_0$ is the cross-section of open tube, $\rho$ represents the (background) air density, $v(x)$ is the velocity of the medium particles at position $x$ in the open tube, $p(x)$ is the pressure deviation from the atmospheric pressure due to the acoustic wave along the open tube. 

In order to express the Lagrangian as a function of the field, the velocity is expressed as the gradient of a "velocity potential": $v(x) = - \partial_x\phi(x) = -\phi'(x)$. The Newton's equation of motion for the air inside the open tube transforms as
\begin{equation}
  \partial_x p = -\dot{v}\rho \qRa
  p = \dot{\phi}\rho,
\end{equation}
and the Lagrangian becomes
\begin{align}
  \cL &= \half \sum_{i=1,2} \rho SL\qty(\dot{\xi}_i^2 - \omega_i^2 \xi_i^2) + \sum_{i=1,2}\int dx\delta(x_i)\dot{\phi}\rho S\xi_i \nn
  &+ \half\int dx S_0\rho\qty[\frac{1}{c^2}\dot{\phi}(x)^2 - \phi'(x)^2].
\end{align}

The corresponding Euler-Lagrange equations give:
\begin{align}
  \ddot{\xi_i} + \omega_i^2 \xi_i &= \dot{\phi}(x_i)/L \qfor i = 1, 2, \\
  \ddot{\phi} - c^2\phi'' &=  - c^2\frac{S}{S_0}\qty[\delta(x_1)\dot{\xi_1} + \delta(x_2)\dot{\xi_2}].
\end{align}
The first equation is the EOM for the Helmholtz resonators, and the second equation is the EOM for the sound wave in the tube. Integrating the second equation over a small region $(x_i^-,x_i^+)$, we obtain the boundary condition at $x = x_{1,2}$
\begin{equation}
  \frac{S}{S_0}\dot{\xi_i} = \phi'(x_i^+) - \phi'(x_i^-) \qRa
  \frac{\rho S}{S_0}\ddot{\xi_i} = p'(x_i^+) - p'(x_i^-).
\end{equation}
In areas $x \neq x_{1,2}$, we have the sound wave equation $\ddot{\phi} -c^2 \phi'' = 0$.
Other boundary conditions can be derived:
\begin{equation}
  \eval{\frac{\partial \cL}{\partial (\partial_x\phi)}}_{x_i} = 0 \qRa
  p(x_i^+) = p(x_i^-),
\end{equation}
corresponding to the continuity of pressure at positions of the resonators $x_{1,2}$.

Eliminating the auxiliary quantity $\phi$, the EOM and boundary conditions are summarized as follows:
\begin{subequations}
  \begin{align}
    p'' - \frac{1}{c^2}\ddot{p} &= 0,\\
    \ddot{\xi_i} + \omega_i^2 \xi_i &= \frac{p(x_i)}{\rho L}, \\
    p'(x_i^+) - p'(x_i^-) &= \frac{\rho S}{S_0}\ddot{\xi_i}, \\ 
    p(x_i^+) - p(x_i^-) &= 0.
  \end{align}
\end{subequations}
The first equation is the acoustic pressure wave equation for an open pipe, the second is the EOM for the Helmholtz resonators, the third corresponds to the flow conservation equation at the connection between the Helmholtz resonator and the open pipe, and the fourth equation is the continuity equation for the acoustic pressure at the connection of the open pipe.

Expand $\xi_i$ and $p(x)$ in terms of Fourier components:
\begin{align*}
    \xi_i &= a_ie^{-i\omega t}, \\
    p(x) &= 
      \begin{cases}
        \qty(c_1^+ e^{iqx} + c_1^- e^{-iqx})e^{-i\omega t} &\quad x \le x_1 \\
        \qty(c_2^+ e^{iqx} + c_2^- e^{-iqx})e^{-i\omega t} &\quad x_1 < x < x_2 \\
        \qty(c_3^+ e^{iqx} + c_3^- e^{-iqx})e^{-i\omega t} &\quad x \ge x_2
      \end{cases},
\end{align*}
with $q = \omega/c$ the wave vector. Due to the open nature of the coupling pipe, no incident waves are injected from either side, $c_1^+ = c_3^- = 0$. With these ansatz solutions, the equations of motion and boundary conditions become (with all $c_i^\pm$ eliminated)
\begin{equation}
  \mqty(
    \omega_1^2 - \omega^2 -2i\kappa\omega & 2\kappa\omega e^{i\theta} \\
    2\kappa\omega e^{i\theta} & \omega_2^2 - \omega^2 -2i\kappa\omega )
  \mqty(a_1 \\ a_2) = 0,
\end{equation}
where 
\begin{equation}
 \kappa = \frac{Sc}{4S_0L} \qand
 \theta = ql - \frac{\pi}{2}
\end{equation}
with $l = x_2 - x_1$. 
By dividing $\omega_1 + \omega_2 \sim 2\omega$, we get:
\begin{equation}
  \mqty(\omega_1 - i\kappa & \kappa e^{i\theta} \\
    \kappa e^{i\theta} & \omega_2 - i\kappa)
  \mqty(a_1 \\ a_2)
  = \omega \mqty(a_1 \\ a_2).
\end{equation}

\section{Magnetic system}
\label{app:magnetic}

For the magnetic system with two YIG spheres placed at $\br_{1,2}$ beside a coplanar wave guide, we assume the size of the sphere is much smaller than the wave length of the electromagnetic waves. Then the EOM for the magnetic moments is the LLG equations
\begin{equation}
  \label{A:LLG}
  \pdv{\mb_i}{t} + \gamma \mb_i\times\qty[\bH_i + \bh(\br_i)] = 0,
\end{equation}
and the EOM for the magnetic field (flux density) is
\begin{equation}
  \label{A:maxwell}
    \frac{1}{c^2}\pdv[2]{t}\bh - \nabla^2 \bh
    = -\frac{1}{c^2}\pdv[2]{t}\mb +\nabla(\nabla\cdot\mb).
\end{equation}



\begin{figure}[t]
  \centering
  \includegraphics[width=0.8\columnwidth]{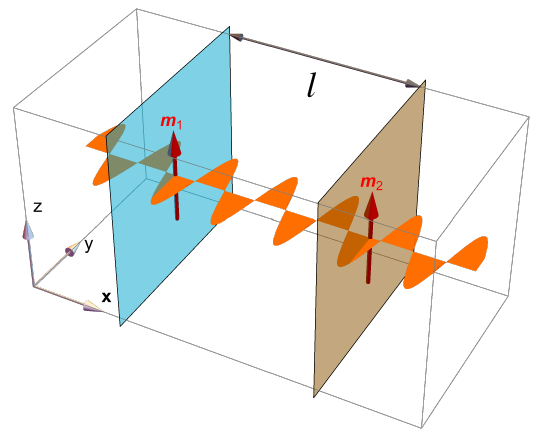}
  \caption{The magnetic moments of two infinitely large films coupled via $y$-polarized electromagnetic waves.}
  \label{fig:mag_diagram}
\end{figure}

We reduce the complexity of two YIG spheres into a quasi-one-dimensional model, represented by two parallel magnetic thin films separated by a distance \( l = x_2 - x_1 \) as shown in \Figure{fig:mag_diagram}. The system exhibits translational symmetry in the \( y \)-\( z \) plane, and the thickness of the thin films \( d \) is significantly smaller than the wavelengths of the electromagnetic waves under consideration, allowing us to treat the films as effectively infinitely thin. The magnetic moments within these films are represented as macrospins, denoted by \( \mb_{1,2} \), with a uniform saturation magnetization \( M_s \) and an equilibrium magnetization aligned in the \( \hbz \) direction. By integrating \Eq{A:maxwell} over the region \( (x_i^-, x_i^+) \) across the magnetic thin films, we derive the connection relation for the magnetic field \( \bh \) on either side of the thin films.
\begin{equation}
    \p_x \bh(x_i^+) - \p_x \bh(x_i^-) = \frac{M_sd}{c^2} \p_t^2 \mb_i.
\end{equation}

In the magnetic system model in the main text, the coplanar waveguide restricts the polarization direction of the magnetic field. Therefore, we only need to consider the coupling mediated by the y-polarized magnetic field here (i.e., consider the coupling mediated by plane waves with wave vectors perpendicular to the plane of the film).
\begin{subequations}
  \begin{align}
    \p_t^2 m_i^y + \omega_i^2 m_i^y &= \omega_i\gamma h_y(x_i) \\
    \p_t^2 h_y - c^2\p_x^2 h_y &= 0, \\ 
    \p_x h_y(x_i^+) - \p_x h_y(x_i^-) &= \frac{M_sd}{c^2} \p_t^2 m_i^y, \\
    h_y(x_i^+) - h_y(x_i^-) &= 0,
  \end{align}
\end{subequations}
where $\omega_i = \gamma H_i $. Similar to the acoustic system, by expanding $h_y$ and $m_y$ in terms of Fourier components and eliminating the Fourier components of $h_y$, we obtain:
\begin{equation}
  \mqty(
    \omega_1 - i\kappa & \kappa e^{i\theta} \\
    \kappa e^{i\theta} & \omega_2 - i\kappa )
  \mqty( a_1 \\ a_2)
  = \omega \mqty(a_1 \\ a_2)
\end{equation}
where $a_i$ are the Fourier components of $m_i^y$ and
\begin{equation}
  \kappa = \frac{qd\omega_\ssf{M}}{4}
  \qand
  \theta = ql + \frac{\pi}{2}.
\end{equation}

\end{document}